\documentclass[aps,pra,twocolumn,amsmath,amssymb]{revtex4-2}
\usepackage{hyperref}
\usepackage{graphicx}

\usepackage{bm}
\usepackage{braket}

\usepackage{booktabs}
\usepackage[dvipsnames]{xcolor}

\renewcommand{\vec}[1]{\mathbf{#1}}

\usepackage{tikz}
\usetikzlibrary{positioning}

\begin{document}

\title{Angular momentum redirection phase of vector beams in a non-planar geometry}
\author{Amy McWilliam}
\author{Claire Marie Cisowski}
\author{Robert Bennett}
\author{Sonja Franke-Arnold}
\affiliation{School of Physics \& Astronomy, University of Glasgow, Glasgow G12 8QQ, United Kingdom}
\date{\today}%

\begin{abstract}
An electric field propagating along a non-planar path can acquire geometric phases. Previously, geometric phases have been linked to spin redirection and independently to spatial mode transformation, resulting in the rotation of polarisation and intensity profiles, respectively. We investigate the non-planar propagation of scalar and vector light fields and demonstrate that polarisation and intensity profiles rotate by the same angle.  The geometric phase acquired is proportional to $j=\ell+\sigma$, where $\ell$ is the topological charge and $\sigma$ is the helicity. Radial and azimuthally polarised beams with $j=0$ are eigenmodes of the system and are not affected by the geometric path. The effects considered here are relevant for systems relying on photonic spin Hall effects, polarisation and vector microscopy, as well as topological optics in communication systems.
\end{abstract}

\maketitle

%~~~~~~~~~~~~~~~~~~~~~~~~~~~~~~~~~~~~~~~~~~~~~~~~~~~~~~~~~~~~~~~~~~~~~~
\section{Introduction}
%~~~~~~~~~~~~~~~~~~~~~~~~~~~~~~~~~~~~~~~~~~~~~~~~~~~~~~~~~~~~~~~~~~~~~~
Throughout history, mirrors have been used for the most sacred and profane purposes, as well as for a multitude of scientific and technological purposes.  The earliest documented mirrors have been constructed in Neolithic times from polished obsidian \cite{ObsidianMirror}, followed by metallic mirrors, not unlike the ones used in this paper, developed in Mesopotamia around 4000 BCE, Venetian molten glass mirrors from around 1400 and the modern day dielectric mirrors.  Contrary to popular misconceptions, mirrors do not ‘swap left and right’ but rather ‘near and far’. In an optical context we might say that the propagation direction of a light beam perpendicular to the mirror surface is reversed.  

Mirror reflection also affects light's (spin and orbital) angular momentum, however in a subtly different way:  the angular momentum component perpendicular to the mirror surface is conserved, whereas the parallel component is reversed – effectively reversing the projection of the angular momentum with respect to the propagation direction. 
When taking light via multiple mirrors along a non-planar trajectory, these successive angular momentum redirections add up, and the propagation of the electric field may be modelled as parallel transport along the beam trajectory \cite{Berry1987a,Bliokh2008}, resulting in a rotation of both the polarisation and intensity profile. 

Polarisation rotation may be understood by considering the action of individual optical elements along the non-planar beam path on the electric field vector \cite{Bortolotti1926,Rytov1938,Vladimirskiy1941}. This has been confirmed using optical fibres curled into a helix \cite{Ross1984, Haldane1986, ChiaoWu1986} or by using a succession of mirror reflections \cite{Kitano1987,BerryNature1987}. 
Propagation along a non-planar trajectory also causes a rotation of the beam intensity profile \cite{MaRamachandran,Patton:19,Galvez1}, which can be seen from simple ray tracing. 

In this paper, we investigate experimentally and theoretically the rotation of intensity, polarisation and vector field profiles, using the same experimental setup to transport a beam of light along a non-planar trajectory. We show systematically that the rotation angle depends solely on the non-planarity of the beam path, and interpret it in terms of geometric phases.  For polarisation rotations, these are the well-established spin-redirection phases: Upon non-planar propagation, the right and left handed circularly polarised components of the light field acquire equal and opposite phases, leading to a rotation of the polarisation ellipse. Similarly, a spatial mode with a given orbital angular momentum (OAM) acquires a phase proportional to its topological charge \cite{Bliokh2006}, which we shall call \emph{orbital-redirection phase}. We show the differential phase shifts acquired by the various modal contributions result in a rotation of the overall intensity profile.  

The geometric phase provides a unifying concept in physics and specifically in optics \cite{Cohen2019,Jisha2021}. It describes a phase modulation that, unlike the dynamic phase, is independent of the optical path length but results exclusively from the geometry of the optical trajectory \cite{Bliokh2015}. Such geometric phases 
play a crucial role in the photonic spin Hall effect \cite{Ling2017,Dai2020,Kim2020} and its scalar equivalent, the acoustic orbital angular momentum Hall effect \cite{Fan2021}, and spin-orbit transformations in general.

In our work we apply the concept of geometric phases to the simplest possible experimental setup, comprising nothing but a succession of metal mirrors in a non-planar configuration. For the first time, however, we study polarisation and image rotation for a wide range of scalar and vector beams, including beams with inhomogeneous spatial polarisation distributions \cite{Zhan:09}, which are non-separable in their spin and orbital degrees of freedom \cite{Souza:14}. 
We systematically show that polarisation profiles and images are rotated by the same angle, confirming the concept of an angular redirection phase as the origin of optical beam rotation.

%~~~~~~~~~~~~~~~~~~~~~~~~~~~~~~~~~~~~~~~~~~~~~~~~~~~~~~~~~~~~~~~~~~~~~~
\section{Experimental setup}
%~~~~~~~~~~~~~~~~~~~~~~~~~~~~~~~~~~~~~~~~~~~~~~~~~~~~~~~~~~~~~~~~~~~~~~

\begin{figure}[h!]
    \includegraphics[width = 0.45\textwidth]{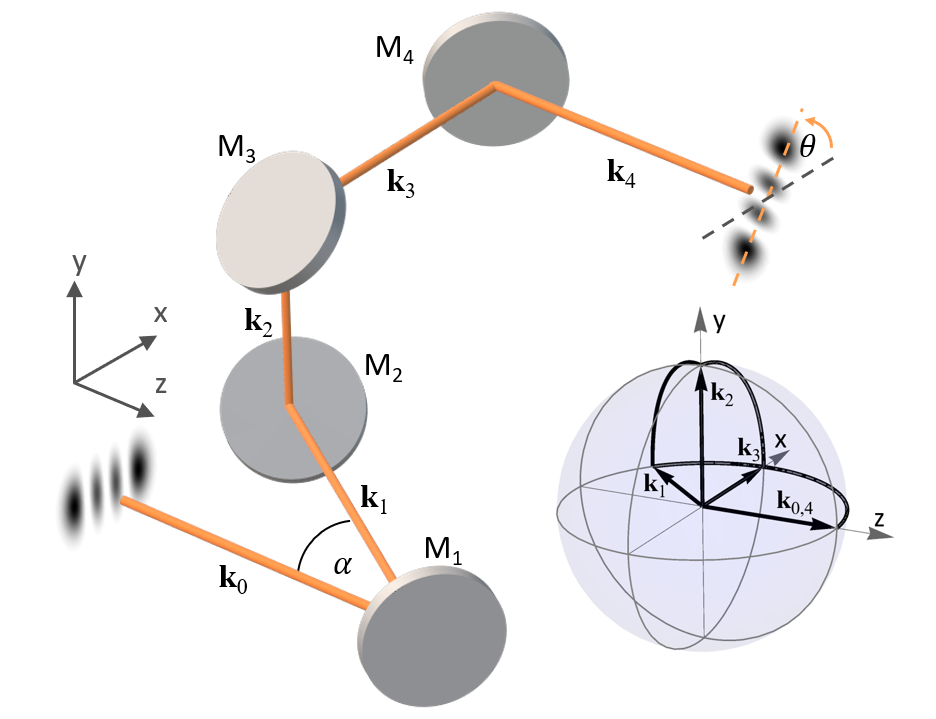}
    \caption{Experimental setup. A succession of mirror reflections takes the beam along a curved non-planar trajectory. $\vec{k}_{n}$ labels the wave vector after the $n$th mirror ($\text{M}_{n}$) reflection. Inset: corresponding path of the wave vector on the $\vec{k}$-sphere.}
    \label{setupFig}
\end{figure}

We generate a non-planar beam path, as outlined in Fig.~\ref{setupFig} and inspired by a setup used in \cite{Chiao1988, Jiao1989}, comprising four mirror reflections.  Mirrors $\text{M}_{2}$, $\text{M}_{3}$ and $\text{M}_{4}$ define a (vertical) plane, and adjusting the position of mirror $\text{M}_{1}$ allows us to access a non-planar geometry.  The beam path is arranged such that the input and output wavevectors $\vec{k}_{0}$ and $\vec{k}_{4}$ point in the same direction (chosen as along the $z$ axis), which allows us to define rotations unambiguously. The rotation of intensity patterns is determined by analysis of images obtained via a CMOS camera, and the (spatially resolved) polarisation rotation by full Stokes tomography \cite{Selyem2019,Rosales-Guzman2020}. 

Experimentally, we parameterize the degree of non-planarity by the angle 
\begin{equation}
\label{eq_theta}
\theta_{\rm NP}= \pi/2-\alpha,
\end{equation}where $\alpha$ is twice the angle of incidence on mirror $\text{M}_{1}$.  Specifically, $\alpha$ is adjusted by shifting the position of mirror $\text{M}_{1}$ and adjusting $\text{M}_{2}$ accordingly. In order to minimise unwanted optical activity we use metallic mirrors.

We generate beam profiles with arbitrary intensity and polarisation structures using a digital micromirror device (DMD), following techniques outlined in \cite{Selyem2019,Rosales-Guzman2020}.
This allows us to investigate homogeneously polarised beams as well as those with spatially varying polarisation profiles, simply by changing the multiplexed hologram pattern on the DMD. While our setup generates vector beams as superpositions of different spatial modes in the horizontal and vertical polarisation components, these may be converted into a modal decomposition of circular polarisations, allowing us to design arbitrary vector modes.

Throughout the paper we are using the coordinate system indicated in Fig.~\ref{setupFig}. A positive rotation angle $\theta$ then corresponds to a clockwise rotation if defined in the direction of beam propagation, which appears anticlockwise when observed on the camera.

%~~~~~~~~~~~~~~~~~~~~~~~~~~~~~~~~~~~~~~~~~~~~~~~~~~~~~~~~~~~~~~~~~~~~~~
\section{Image Rotation}
%~~~~~~~~~~~~~~~~~~~~~~~~~~~~~~~~~~~~~~~~~~~~~~~~~~~~~~~~~~~~~~~~~~~~~~

Spatially resolved ray propagation, taking into account the action and alignment of the various optical components, reveals that any intensity profile exiting the experiment is rotated by the angle $\theta$ defined in (\ref{eq_theta}), as illustrated in Fig.~\ref{setupFig} for the Hermite-Gaussian (HG) mode $\text{HG}_{3,0}$. 

Any mode rotation can be interpreted in terms of geometric phases.  If $\Psi(r,\phi)$ is the original mode expressed in cylindrical coordinates, then a rotation by $\theta$ will result in the mode $\Psi'(r,\phi)=\Psi(r,\phi+\theta).$ Here we define a clockwise rotation around the propagation axis as positive. Expressing this mode in Laguerre-Gaussian (LG) basis, with ${\rm LG}_p^\ell=u_p^\ell(r) \exp(i \ell \phi),$ we can write the original mode as 
\begin{equation}
    \psi(r,\phi)= \sum_{p,\ell}\braket{\text{LG}_{p}^{\ell}|\psi(r,\phi)}   \text{LG}_{p}^{\ell},
\end{equation}
where $\braket{a|b}$ denotes the inner product which may be evaluated as mode overlap.  The rotated mode is then
\begin{align}
\label{eq_decomposition}
\psi'(r,\phi)&= \sum_{p,\ell}\braket{\text{LG}_{p}^{\ell}|\psi(r,\phi+\theta)}   \text{LG}_{p}^{\ell} \nonumber\\
& = \sum_{p,\ell}{\rm e}^{-i \ell \theta}\braket{\text{LG}_{p}^{\ell}|\Psi(r,\phi)}   \text{LG}_{p}^{\ell},  \end{align}
where we have evaluated the inner product as 
\begin{align*}
\braket{\text{LG}_{p}^{\ell}|\Psi(r,\phi+\theta)} & =\braket{u_p^\ell(r) {\rm e}^{i \ell \phi}|\Psi(r,\phi+\theta)} \\
& =\braket{u_p^\ell(r) {\rm e}^{i \ell (\phi-\theta)}|\Psi(r,\phi)} \\
& = {\rm e}^{-i \ell \theta} \sum_{p,\ell} \braket{\text{LG}_{p}^{\ell}|\Psi(r,\phi)}.
\end{align*}
where the second line results from relabelling the angular variables. We find that a rotation by $\theta$ is associated with an orbital-redirection phase of $-\ell \theta$ for every LG mode of topological charge $\ell$. This is of course a direct consequence of the fact that LG modes are eigenmodes of the angular momentum operator $-i \partial_\theta$, and the fact that the angular momentum operator is the generating function of rotations. 

\begin{figure}[ht]
    \includegraphics[width = 0.3\textwidth]{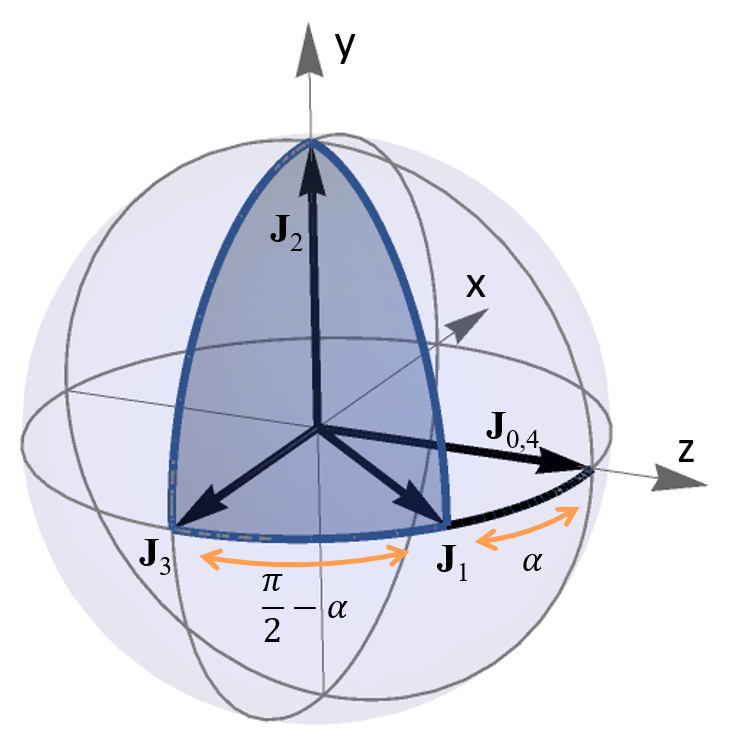}
    \caption{Angular momentum redirection sphere for $\vec{J} = \vec{L}+\vec{S}$, illustrated for the case where $\vec{J}$ is aligned with the initial wave vector.
    }
    \label{fig:AMsphere}
\end{figure}

Redirection-phases are said to be geometric because they depend on the geometry of the path formed on the relevant sphere of angular momentum directions. For paraxial beams, the  (spin/orbital) angular momentum vectors are aligned with the $\vec{k}$ vectors, but the component of $\textbf{S}$ and $\textbf{L}$ perpendicular to the mirror does not change orientation after every mirror reflection \cite{barnett2012}.  
With this in mind, the $\vec{k}$ redirection sphere, shown as the inset in Fig.~\ref{setupFig}, can be translated into an angular momentum redirection sphere, shown in Fig.~\ref{fig:AMsphere}.  For image rotations considered here, $\theta$ is equal to the solid angle $\Omega$, corresponding to the blue curve on  the orbital redirection sphere upon propagation, in Fig.~\ref{fig:AMsphere}. 

Any transverse spatial mode of mode order $N$ can be expressed as superposition of $N+1$ modes from a different mode family with the same mode order. For HG modes $\textrm{HG}_{n,m}$ and LG modes $\textrm{LG}_p^\ell$ the mode number is given by $N=n+m=2p+| \ell |$ respectively. 

This allows us to illustrate the orbital-redirection phase for simple examples:  The first order mode $\text{HG}_{1,0}$ rotated by an angle $\theta$ in the clockwise direction (with beam propagation), may be expressed as
\begin{align}
\text{HG}_{1,0}' & =\frac{1}{\sqrt{2}}(\text{LG}_{0}^{+1} e^{-i\theta}+\text{LG}_{0}^{-1} e^{i\theta}) \nonumber\\
& =\cos\theta \; \text{HG}_{1,0}+\sin\theta \; \text{HG}_{0,1},
\end{align}
in the LG basis and the HG basis respectively.

Similarly, the second order mode $\text{HG}_{2,0}$ mode rotated by $\theta$ can be written as: 
\begin{align}\label{secondOrderHG}
\text{HG}_{2,0}' & =\frac{1}{2}(\text{LG}_{0}^{+2}e^{-i2\theta}-\sqrt{2}\text{LG}_{1}^{0}+\text{LG}_{0}^{-2}e^{+i2\theta}) \nonumber \\ 
& = \cos^2(\theta)\text{HG}_{2,0} +\frac{\text{sin}(2\theta)}{\sqrt{2}}\text{HG}_{1,1} +\sin^{2}(\theta)\text{HG}_{0,2}.    
\end{align}

\begin{figure}[ht]
    \centering
    \includegraphics[width = .9\columnwidth]{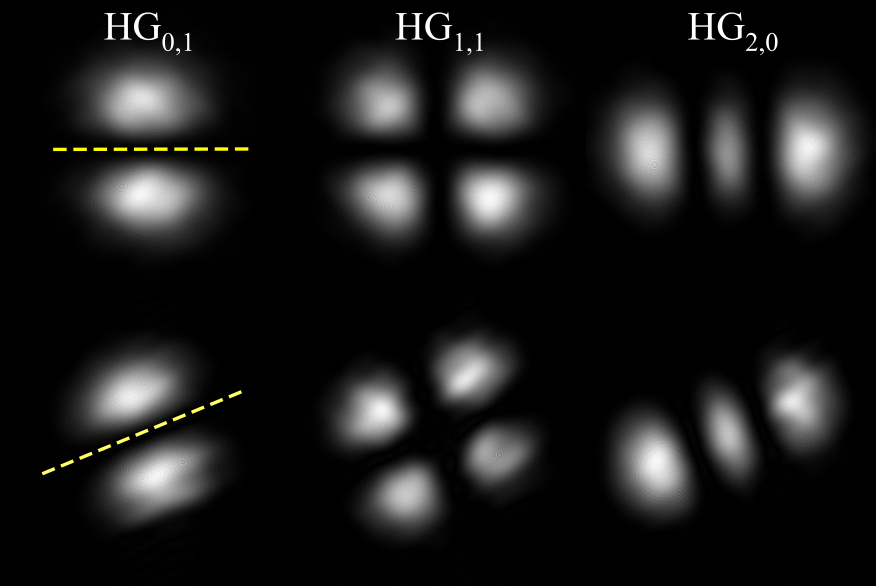}
    \caption{Image rotation: Experimental images (from the camera's perspective) of the original (top) and rotated (bottom row) mode profiles for a fixed non-planar trajectory with $\theta_\mathrm{NP} = (21.4 \pm 0.5)^\circ$, indicated as yellow dashed line.}
    \label{fig:IntensityRotationExamples}
\end{figure}

Measurements of various HG intensity profiles before and after non-planar propagation are shown in Fig.~\ref{fig:IntensityRotationExamples} for a fixed angle $\alpha$.

In order to confirm the relationship between non-planarity and image rotation experimentally, we realise non-planar trajectories, with $\alpha$ set to various angles between $15^\circ$ and $70^\circ$, and identify the image rotation for all 14 ${\rm HG}_\mathrm{n,m}$ modes,  with a mode number $N=n+m \leq 4$.
We first low-pass Fourier filter the camera images to remove artefacts due to diffraction, and then convert them to polar plots. We obtain the rotation angle from the angular offset between the rotated and input beam profiles. 

\begin{figure}[ht]
    \includegraphics[width = 0.45\textwidth]{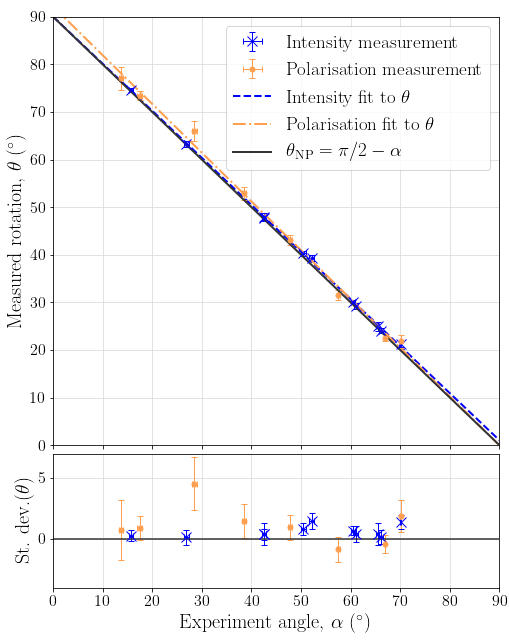}
    \caption{Image and polarisation rotation for non-planar trajectories: Experimental measurements and theoretical prediction of the relationship between non-planarity set by $\alpha$ and the intensity rotation $\theta$. The lower plot shows the difference between the measured $\theta$ and the theoretically expected $\theta_\text{NP}$. The error in $\alpha$ is estimated to be smaller than $0.5^\circ$.
    }
    \label{fig:IntensityRotationPlot}
\end{figure}

As expected, the rotation angle does not depend on the specific transverse input mode. For each geometric configuration, determined by angle $\alpha$, we therefore evaluate the rotation $\theta$ by averaging over the individual rotation angles found for our 14 test modes. Fig.~\ref{fig:IntensityRotationPlot} shows the rotation angles $\theta$ of the intensity distribution for various incidence angles $\alpha$, confirming that $\theta=\theta_{\rm NP}=\pi/2 - \alpha$, i.e.~the image rotation angle is directly given by the degree of non-planarity of the beam trajectory.

%~~~~~~~~~~~~~~~~~~~~~~~~~~~~~~~~~~~~~~~~~~~~~~~~~~~~~~~~~~~~~~~~~~~~~~
\section{Polarisation rotation}\label{PolSection}
%~~~~~~~~~~~~~~~~~~~~~~~~~~~~~~~~~~~~~~~~~~~~~~~~~~~~~~~~~~~~~~~~~~~~~~

Using the same experimental setup we confirm that propagation along non-planar trajectories rotates not only images, but also the axis of the polarisation ellipse characterizing the electric field. We study the rotation of homogeneously polarised beams, independent on the spatial state of the mode, here illustrated for fundamental Gaussian beams.

The propagation of the light along a non-planar trajectory can be modelled using Jones vector formalism \cite{Jones:41}, where the action of each mirror on the polarisation vector is described by a Jones matrix that arises from simple ray optics. As we are working in a 3D geometry, the electric field is described by 3D Jones vectors $\mathbf{E}=E_h \hat{h} + E_{v}\hat{v}+E_{z}\hat{z},$ where $\hat{h}$, $\hat{v}$ and $\hat{z}$ are unit vectors in the horizontal, vertical and propagation direction.  

As shown in Appendix \ref{JonesAppendix}, the action of propagation along the non-planar trajectory is described by the Jones matrix,
\begin{equation}\label{PMatrix}
P = \left(\begin{array}{ccc}
    \cos\theta & -\sin\theta & 0 \\
    \sin\theta & \cos\theta & 0\\
    0 & 0 & 1\\
\end{array}\right)
\end{equation}
where, $\theta = \theta_\mathrm{NP}=\pi/2 - \alpha$. 
This is, of course, a general rotation matrix, showing that the mirror system rotates an input polarisation state by an angle $\theta$ about the $z$ axis.

Similarly to image rotation, a polarisation rotation can be explained in terms of geometric phases, in this case between the circular polarisation components. 
Any paraxial linearly polarized beam of light, propagating along the $z$ direction, may be written as:
\begin{equation}
\textbf{E}= E_0(r,\phi) \left(\cos \theta_0 \hat{h}+ \sin \theta_0 \hat{v} \right), 
\end{equation}
where $E_0(r,\phi)$ describes the transverse spatial mode, and $0\le\theta_0\le 2\pi$ is the angle of the polarisation direction to the horizontal. 
Applying the rotation matrix \eqref{PMatrix} results in the rotated mode 
\begin{equation}
\textbf{E}'= E_0(r,\phi) \left(\cos (\theta_0+\theta) \hat{h}+ \sin (\theta_0+\theta) \hat{v} \right). 
\end{equation}
One may easily verify that the initial and rotated field, expressed instead in terms of the circular polarisation basis $\hat{r}=(\hat{h}-i\hat{v})/\sqrt{2}$ and $\hat{l}=(\hat{h}+i\hat{v})/\sqrt{2}$, become 
\begin{align}
\textbf{E} & = (E_0(r,\phi)/\sqrt{2}) \left(\mathrm{e}^{i \theta_0} \hat{r}+ \mathrm{e}^{-i \theta_0} \hat{l} \right), \\
\textbf{E}' & = (E_0(r,\phi)/\sqrt{2}) \left(\mathrm{e}^{i (\theta_0+\theta)} \hat{r}+ \mathrm{e}^{-i (\theta_0+\theta)} \hat{l} \right). 
\end{align}
We can interpret the rotation as the right and left hand polarisation component, carrying an angular momentum of $-\hbar$ and $+\hbar$ per photon, acquire a phase factor of $\exp(i\theta)$ and $\exp(-i\theta)$, respectively. This result also holds for general elliptical beams, but for simplicity we have omitted the lengthy calculation.

A rotation by $\theta$ is associated with a spin-redirection phase of $-\sigma\theta$ for every circularly polarised modes of helicity $\sigma$, leading to polarisation rotation, and an orbital-redirection phase of $-\ell \theta$ for every spatial mode component with an OAM of $\ell \hbar$. For both cases, the rotation angle $\theta$ is equal to the solid angle $\Omega$ formed by the path traced on the appropriate (spin or orbital) angular momentum redirection sphere shown in Fig.~\ref{fig:AMsphere}.

\begin{figure}[t]
  \centering
    \includegraphics[width=0.9\columnwidth]{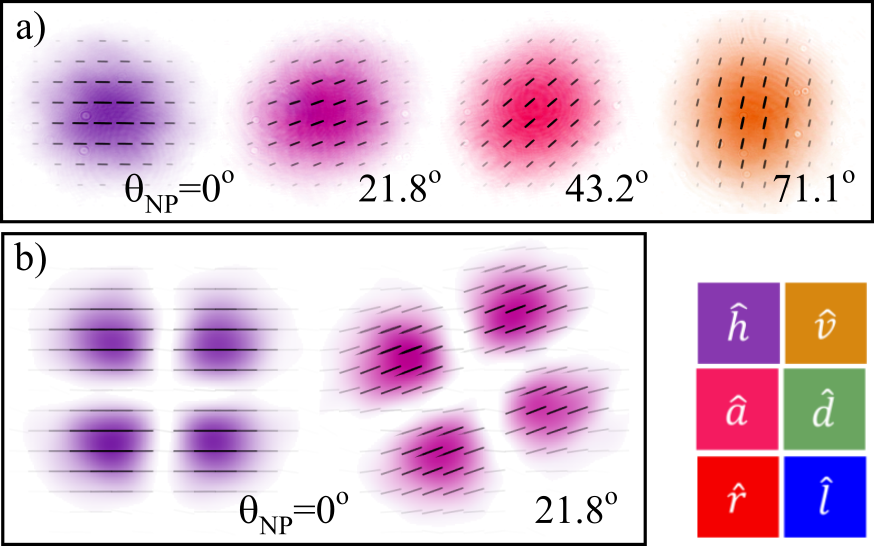}
  \caption{Polarisation rotation: (a) Experimental observation of polarisation rotation of a horizontally polarised Gaussian beam (left), for different non-planar geometries  
  (b) Polarisation profile of a homogeneously polarised $\rm{HG}_{1,1}$ mode before and after rotation.
 The polarisation colour scheme used throughout is shown as inset, with intensity represented as opacity.}
 \label{fig:HorizontalRotation}
\end{figure}

We confirm this experimentally, by identifying the polarisation rotation of an initially horizontally polarised beam after it is passed along a non-planar trajectory with various angles of $\theta_\mathrm{NP}=\pi/2-\alpha$. A selection of these measurements are shown in Fig.~\ref{fig:HorizontalRotation}. 

Polarisation rotation of homogeneously polarised light is independent on the spatial mode, as illustrated in Fig.~\ref{fig:HorizontalRotation}(b) on the example of a horizontally polarised $\rm{HG}_{1,1}$ mode. It is evident that both the intensity and polarisation profiles have rotated by the same angle. 

Our spatially resolved measurements show that, as expected, the polarisation remains homogeneous across the beam profile. We determine the rotation angle $\theta$ by averaging over the spatially resolved Stokes measurements before and after propagation, with the error in $\theta$ given by the standard deviation. A plot of measured experimental angle, $\alpha$ against measured polarisation rotation, $\theta$, is shown in Fig.~\ref{fig:IntensityRotationPlot}, confirming that the polarisation rotation is equal to the non-planarity parameter, $\theta=\theta_\mathrm{NP} = \pi/2 - \alpha$.

%~~~~~~~~~~~~~~~~~~~~~~~~~~~~~~~~~~~~~~~~~~~~~~~~~~~~~~~~~~~~~~~~~~~~~~
\section{Rotation of vector beams}
%~~~~~~~~~~~~~~~~~~~~~~~~~~~~~~~~~~~~~~~~~~~~~~~~~~~~~~~~~~~~~~~~~~~~~~
\begin{figure*}[ht]
 \centering
 \includegraphics[width=1\linewidth]{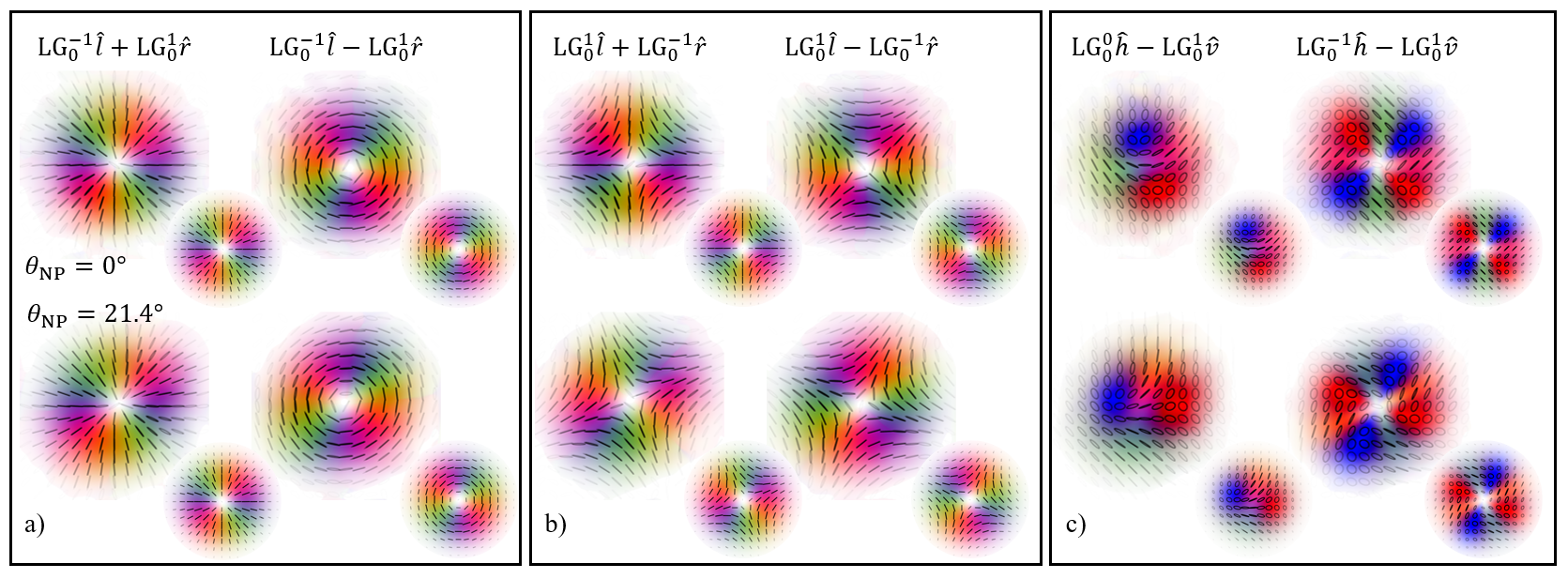}
 \caption{Polarisation plots of experimentally measured vector beams before (top row) and after (bottom row) the system, for a fixed rotation angle of $\theta = (21.4\pm0.5)^\circ$. Beams in (a, b) contain only linear polarisation and are of the form ${\rm{LG}}_0^{\pm1}\hat{l}+\exp{(-i \varphi)} \rm{LG}_0^{\pm1}\hat{r}$, for $\varphi = (0, \pi)$.
 The first column in (c) is a Poincar\'e beam and the second column shows a beam containing polarisations along a great circle on the Poincar\'e sphere. Smaller inserts show the corresponding theory plots.}
 \label{fig:VectorBeamRotation}
\end{figure*}
Finally, we study the rotation of beams of light presenting an inhomogeneous polarisation distribution. These are non-separable in spatial and polarisation degrees of freedom. General vector beams can be written as $\bm{\psi}=\psi_\mathrm{l}(r,\varphi) \hat{l}+ \psi_\mathrm{r}(r,\varphi) \hat{r},$ but here we restrict ourselves, without loss of generality, to beams of the form
\begin{equation} 
\label{eq_vb}
\bm{\psi}={\rm LG}_{p_1}^{\ell_1} \hat{l}+ e^{-i\varphi} {\rm LG}_{p_2}^{\ell_2} \hat{r}.
\end{equation}
As discussed in the previous sections, the action of the non-planar propagation results in orbital redirection phase of $\exp(-i\ell\phi)$ for mode with OAM of $\ell \hbar$, and a spin redirection phase of $\exp(-i\sigma\phi)$ for a mode with a spin of $\sigma \hbar$ along the propagation direction. The total geometric phase is hence proportional to the total \emph{total} angular momentum number $j = \ell+ \sigma$.  Applied to our vector beam Eq.(\ref{eq_vb}), this leads to a simultaneous rotation of the 
intensity pattern and the polarisation by $\theta$:
\begin{equation}
\bm{\psi}' = {\rm e}^{-i(\ell_1+1)\theta}{\rm LG}_{p_1}^{\ell_1} \hat{l}
+ {\rm e}^{i\varphi}{\rm e}^{-i(\ell_2-1)\theta} {\rm LG}_{p_2}^{\ell_2} \hat{r}.
\end{equation}
In particular, if $j = 0$, the original beam will be recovered after rotation, meaning that the output beam will be indistinguishable from the input, both in polarisation and intensity. As $\sigma$ can only take values $\pm 1$, this only happens for LG modes with $l_1=-1$ and $l_2=1$. These beams remain unchanged and can be considered the eigenmodes of the system and therefore, can be thought of as conserved quantities.
This makes intuitive sense, as beams with these properties are rotationally symmetric, both in their intensity and polarisation profile, making them rotation invariant.  For the simple case of $p_{1,2} = 0$, we obtain radial polarisation for $\varphi = 0$, and azimuthal polarisation for $\varphi=\pi$, as shown in Fig.~\ref{fig:VectorBeamRotation}(a).
However, for all other values of $j$, the angular redirection phase will cause a rotation of the beam.  Fig.~\ref{fig:VectorBeamRotation}(b, c), shows the rotation of a selection of general vectorial beams for which $j\neq0$, including counter-rotating vector beams, for which $\ell_1 = 1$ and $\ell_2 = -1$, these beams correspond to the mirror images of radially and azimuthally polarized beams.

%~~~~~~~~~~~~~~~~~~~~~~~~~~~~~~~~~~~~~~~~~~~~~~~~~~~~~~~~~~~~~~~~~~~~~~
\section{Conclusion}
%~~~~~~~~~~~~~~~~~~~~~~~~~~~~~~~~~~~~~~~~~~~~~~~~~~~~~~~~~~~~~~~~~~~~~~

The orbital and spin angular momentum refer to fundamentally different properties of light: The former is extrinsic and arises from twisted phasefronts of a light beam or photon wavefunction, the latter is intrinsic, relates to the vector nature of electromagnetic fields or more specifically their circular polarisation.  It is therefore surprising that taking a light beam along a non-planar trajectory affects both spin components in the same way, by adding an angular momentum redirection phase of $\exp(-i j \theta_\mathrm{NP})$, where $j=\sigma+\ell$ is the combined spin and angular momentum number of the light, and  $\theta_\mathrm{NP}$ parametrizes the non-planarity of the trajectory. 

We have experimentally and theoretically looked at the rotation of intensity, polarisation and vector field profiles, that arises from propagation along a closed non-planar trajectory. This rotation has been interpreted in terms of both spin and orbital redirection phases, for the polarisation and intensity, respectively. We have demonstrated that the  intensity and polarisation rotate in the same direction by the same angle, which is dependent on the solid angle enclosed by the path on the associated redirection sphere. By looking at the rotation of vector modes, with spatially varying polarisation, we have shown that the geometric phase acquired is proportional to the total angular momentum number of the beam, $j = \ell+ \sigma$. When $j = 0$, the total geometric phase is null, and the polarisation profile appears unrotated by our system. This is the case for radial and azimuthally polarised modes, which can be considered as eigenmodes of the system. 

Due to their robustness against aplanarity, radial and azimuthal vector modes constitute ideal candidates for optical communication based on optical fibres, inherently subject to twisting and bending constraints \cite{MaRamachandran},
and for microscopy applications, where these modes can also be tightly focused \cite{Youngworth:00}. 
Our findings are also relevant for twisted optical cavities and mirror-based resonators \cite{Clark2020,Habraken2007, Forbes2019} and for quantum metrology \cite{DAmbrosio2013}. 
The redirection phases evidenced in this work can be linked to optical Hall effects, as both can be attributed to the existence of Berry potentials \cite{Bliokh2006}, and also constitute elegant additions to the landscape of geometric phases of light, so far dominated by their planar counterparts \cite{Galvez2003,Milione2011}.

\appendix

\section{Basis rotation}\label{BasisAppendix}

We here give an explicit derivation of image rotation in terms of a basis change to the LG mode basis, linking the rotation angle $\theta$ and the orbital redirection phase. We start with the general case and then give the transformation matrices for specific examples of some HG modes used in the experiment.

An HG mode rotated by an angle $\theta$, denoted $\ket{\text{HG}'}$, can be written as a superposition of fundamental HG modes by first expressing the mode in the Laguerre-Gaussian (LG) basis, applying a rotation matrix $\text{R}_{\text{LG}}(\theta)$, then returning to the HG basis \cite{ONEIL200035}:
\begin{equation}\label{general}
\ket{\text{HG}'}=\text{B}_{\text{HG}\leftarrow\text{LG}}\cdot\text{R}_{\text{LG}}(\theta)\cdot\text{B}_{\text{LG}\leftarrow\text{HG}}\ket{\text{HG}} \end{equation}
where $\text{R}_{\text{LG}}(\theta)=\text{diag}(e^{-i\ell_1\theta},e^{-i\ell_{2}\theta},..,e^{-i\ell_{N+1}\theta})$ imparts an orbital-redirection phase, $e^{-i\ell\theta}$, to the LG component of topological charge $\ell$. Since LG and HG modes both form their own complete and orthonormal basis, one can be converted into a superposition of the other via:
\begin{align}
\ket{\text{HG}_{n,m}}&=\sum_{\ell,p}\braket{\text{LG}_{p}^{\ell}|\text{HG}_{n,m}}   \ket{\text{LG}_{p}^{\ell}}\\
 \ket{\text{LG}_{p}^{\ell}}&=\sum_{n,m}\braket{\text{HG}_{n,m}|\text{LG}_{p}^{l}}\ket{\text{HG}_{n,m}}
\end{align}
where the overlaps $\braket{\text{LG}_{p}^{l}|\text{HG}_{n,m}}$ and $\braket{\text{HG}_{n,m}|\text{LG}_{p}^{l}}$ yield the elements of the mode conversion matrices. While complicated general expressions for these overlaps exist in the literature, it is simplest to calculate these directly from the overlap of the respective modes using:
\begin{equation}  \braket{a|b}=\iint (u_{a}(x,y,0))^{*}u_{b}(x,y,0)\text{d}x\text{d}y \label{overlapIntegral} \end{equation} 
where $u$ are the amplitudes of the beams.

For first order modes, the explicit forms of the matrices describing the mode transformations are
\begin{align}
 &\text{B}_{\text{HG}\leftarrow \text{LG}}^{\text{N}=1}=\frac{1}{\sqrt{2}}\left(\begin{array}{cc}
    1 & 1 \\
    i & -i
\end{array}\right)\\\notag \\
& \text{R}^{\text{N}=1}_{\text{LG}}(\theta)=
\left(\begin{array}{cc}
    e^{-i\theta} & 0\\
    0 &  e^{+i\theta} \\
\end{array}\right)\\\notag \\
& \text{B}_{\text{LG}\leftarrow \text{HG}}^{\text{N}=1} =\frac{1}{\sqrt{2}}\left(\begin{array}{cc}
   1& -i\\
    1 & i
\end{array}\right)
\end{align}

For second order modes, the transformations are:
\begin{align}
 &\text{B}_{\text{HG}\leftarrow \text{LG}}^{\text{N}=2}=\left(\begin{array}{ccc}
    \frac{1}{2}& -\frac{1}{\sqrt{2}} & \frac{1}{2}\\
    \frac{i}{\sqrt{2}} & 0 & -\frac{i}{\sqrt{2}} \\
    -\frac{1}{2} & -\frac{1}{\sqrt{2}} &-\frac{1}{2} 
\end{array}\right)\label{secondOrderBLGtoHGMatrix}\\\notag \\
& \text{R}^{\text{N}=2}_{\text{LG}}(\theta)=
\left(\begin{array}{ccc}
    e^{-i2\theta} & 0 & 0\\
    0 & 1 & 0 \\
    0 & 0 & e^{+i2\theta}
\end{array}\right) \\\notag \\
& \text{B}_{\text{LG}\leftarrow \text{HG}}^{\text{N}=2} =\left(\begin{array}{ccc}
    \frac{1}{2} & -\frac{i}{\sqrt{2}} &-\frac{1}{2}\\
    -\frac{1}{\sqrt{2}} & 0 & -\frac{1}{\sqrt{2}} \\
    \frac{1}{2} & \frac{i}{\sqrt{2}} & -\frac{1}{2}
\end{array}\right) 
\end{align}

%~~~~~~~~~~~~~~~~~~~~~~~~~~~~~~~~~~~~~~~~~~~~~~~~~~~~~~~~~~~~~~~~~~~~~~
\section{Polarisation rotation}\label{JonesAppendix}
%~~~~~~~~~~~~~~~~~~~~~~~~~~~~~~~~~~~~~~~~~~~~~~~~~~~~~~~~~~~~~~~~~~~~~~

For this, we need to consider a system of three orthogonal vectors corresponding to each mirror, the incident wavevector, $\vec{k}$, and two orthogonal polarisation components, one which lies in the plane of incidence of the mirror, $\vec{p}$, and one which lies perpendicular to it, $\vec{s}$.
In this appendix we apply the three-dimensional polarisation ray tracing approach discussed in \cite{yun_three-dimensional_2011} to our system. The action of an optical element $q$ on a polarisation vector ${\vec{e}}_q$ is;
\begin{equation}\label{globalEq}
    {\vec{e}}_q = P_q \cdot {\vec{e}}_{q-1}
\end{equation}
where $P_q$ is some $3\times 3$ matrix to be found. This equation is written in \emph{global} coordinates. It is equivalent to another equation written in \emph{local} coordinates, namely;
\begin{equation}
    w_q = J_q\cdot w_{q-1}
\end{equation}
Any general polarisation vector can be written in terms of a pair of orthogonal basis states $a$ and $b$. Then the global relation \eqref{globalEq} becomes the following pair of statements;
\begin{align}
    {\vec{e}}'_a &= P_q \cdot {\vec{e}}_{a} & {\vec{e}}'_b &= P_q \cdot \vec{e}_{b}
\end{align}
which represent six equations in total. The matrix $P_q$ has nine elements, so we need three more equations to uniquely define it. We choose the matrix form of the law of reflection (in global coordinates)
\begin{equation}
    {\vec{k}}_q = P_q \cdot \vec{k}_{q-1}
\end{equation}
furnishing us with nine equations in nine unknowns, which can be solved for the elements of $P_q$. The properties of each optical element are specified by its Jones matrix, which works on \emph{local} Jones vectors. We therefore need to transform from global to local coordinates, let the optical element act, then transform back to global coordinates. All of this should be contained within $P_q$, so it will have the form;
\begin{equation}\label{PqMatrix}
    P_q = O_{G \leftarrow L} \cdot J_q \cdot O_{L \leftarrow G}
\end{equation}
where $O_{G \leftarrow L}$ transforms from local to global and $O_{L \leftarrow G}$ transforms from global to local. 

We will first consider the global to local transformation. Our orthogonal coordinate system before the optical element is $\{{\vec{s}}_q,{\vec{p}}_q,{\vec{k}}_{q-1}\}$, which we need to project onto the local $z$ direction for that element. This kind of change of basis is obtained by a matrix whose rows are the coordinate vectors of the new basis vectors (local, $\{{\vec{s}}_q,{\vec{p}}_q,{\vec{k}}_{q-1}\}$) in the old basis (global, $\{ {\vec{x}}, {\vec{y}}, {\vec{z}}\} $), so; 
\begin{equation}\label{GtoL}
  O_{L \leftarrow G} =   \left(\begin{array}{ccc}
    s_{x,q} & s_{y,q} & s_{z,q-1} \\
  p_{x,q} & p_{y,q} & p_{z,q-1}\\
  k_{x,q} & k_{y,q} & k_{z,q-1}\\
\end{array}\right)
\end{equation}
To transform from local to global coordinates, we take the inverse of this but replace the basis vectors with those \emph{after} the element. 
\begin{equation}\label{LtoG}
  O^{-1}_{G \leftarrow L} =    O^{T}_{G \leftarrow L} =  \left(\begin{array}{ccc}
    s'_{x,q} & p'_{x,q} & k_{x,q} \\
  s'_{y,q} & p'_{y,q} & k_{y,q}\\
  s'_{z,q} & p'_{z,q} & k_{z,q}\\
\end{array}\right)
\end{equation}
where the first equality follows from the fact that this is an orthogonal matrix, since the basis vectors $\{{\vec{s}}_q,{\vec{p}}_q,{\vec{k}}_{q-1}\}$ are orthogonal. This now means the matrix $P_q$ shown in Eq.~\eqref{globalEq} is fully determined if the Jones matrix of the particular optical element and the vectors $\{{\vec{s}}_q,{\vec{p}}_q,{\vec{k}}_{q-1}\}$ are known. These can all be expressed entirely in terms of ${\vec{k}}_q$ and ${\vec{k}}_{q-1}$;
\begin{align}\label{sAndpEq}
    {\vec{s}}_q &= \frac{{\vec{k}}_{q-1} \times {\vec{k}}_q}{|{\vec{k}}_{q-1} \times {\vec{k}}_q|} & {\vec{s}}'_q &= {\vec{s}}_q \notag \\
    {\vec{p}}_q &= {\vec{k}}_{q-1} \times {\vec{s}}_q & {\vec{p}}'_q &= {\vec{k}}_{q} \times {\vec{s}}'_q
\end{align}
The power of this method lies in the fact that $Q$ optical elements can be cascaded by multiplying their $P_q$ matrices to find an overall matrix $P$;
\begin{equation}
    P = P_Q \cdot P_{Q-1}\cdot \ldots\cdot P_q\cdot \ldots \cdot P_2 \cdot P_1
\end{equation}
Therefore to describe a system all we need are the Jones matrices of each element and the $\vec{k}$ vectors, which in turn will be determined by the position of each mirror. 

As shown in Fig.~\ref{setupFig} in the main text, we have four mirrors and five $\vec{k}$ vectors. The Jones matrix for reflection at an ideal mirror is;
\begin{equation}\label{mirrorJones}
  J_q =  \left(\begin{array}{ccc}
    1 & 0 & 0 \\
  0 & -1 & 0\\
  0 & 0 & 1\\
\end{array}\right)
\end{equation}
and the five $\vec{k}$ vectors are;
\begin{align}\label{kVecList}
    {\vec{k}}_0 &= \begin{pmatrix} 0 \\ 0 \\ 1 \end{pmatrix} & {\vec{k}}_1 &= \begin{pmatrix} \sin \alpha \\ 0 \\ -\cos \alpha \end{pmatrix} & {\vec{k}}_2 &= \begin{pmatrix} 0 \\ 1 \\ 0 \end{pmatrix} \notag \\
    {\vec{k}}_3 &= \begin{pmatrix} 1 \\ 0 \\ 0 \end{pmatrix} & {\vec{k}}_4 &= \begin{pmatrix} 0 \\ 0 \\ 1 \end{pmatrix}={\vec{k}}_0
\end{align}
Using these in Eqs.~\eqref{PqMatrix}-\eqref{mirrorJones}, the combined matrix $P_q$ for this system turns out to be;
\begin{equation}
    P = P_4 \cdot P_3 \cdot P_2 \cdot P_1 = \left(\begin{array}{ccc}
    \sin \alpha & -\cos \alpha & 0 \\
  \cos \alpha & \sin \alpha & 0\\
  0 & 0 & 1\\
\end{array}\right)
\end{equation}
Defining $\theta = \pi/2 - \alpha$ as in the main text, we end up with;
\begin{equation}
    P = P_4 \cdot P_3 \cdot P_2 \cdot P_1 =  \left(\begin{array}{ccc}
    \cos \theta & -\sin \theta & 0 \\
  \sin \theta & \cos \theta & 0\\
  0 & 0 & 1\\
\end{array}\right)
\end{equation}
which is Eq.~\eqref{PMatrix} in the main text. This represents an anticlockwise rotation of points in the $xy$ plane around the $z$ axis. Since the beam is travelling in the positive $z$ direction, this represents a \emph{clockwise} rotation when viewed along the propagation axis of the beam.

\section{Expressions for LG and HG modes}\label{beamsAppendix}

The Laguerre-Gauss modes are expressed in cylindrical coordinates $r = \sqrt{x^2+y^2}$, $\phi = \arctan(y/x)$, and are given by;
\begin{equation}
    \textrm{LG}^\ell_p = \frac{C^\ell_p}{w} \left[\frac{r \sqrt{2}}{w} \right]^{|\ell|}L^{|\ell|}_p\left(\frac{2 r^2}{w^2}\right) e^{-\frac{r^2}{w^2} +  i\left(k \frac{r^2 z}{2(z_\mathrm{R}^2+z^2)}+\ell \phi+ \Phi\right)}
\end{equation}
where $C^\ell_p = \sqrt{\frac{2 p!}{\pi  (| \ell| +p)!}}$, $\Phi = -(2p +|\ell|+1)\chi$ with $\chi =\arctan(z/z_\textrm{R})$ is the Gouy phase, $L^{\vert \ell\vert}_p$ is an associated Laguerre polynomial, $w=w_0 (1+\left(z/z_\mathrm{R}\right)^2)^{1/2}$ is the beam radius for waist $w_0$ and  $z_\textrm{R}=\pi w^2_0/\lambda$ is the Rayleigh range. The Hermite-Gauss modes are given by 
\begin{equation}
    \textrm{HG}_{nm} = \textrm{HG}_{n}(x)\textrm{HG}_{m}(y)
\end{equation}
where
\begin{align}
   \textrm{HG}_{n}(x) =  \frac{C_{n}}{\sqrt{w}}H_n\left(\frac{\sqrt{2} x}{w}\right)e^{-\frac{x^2}{w^2}+\frac{i k x^2 z}{2 (z^2+z_\mathrm{R}^2)}-i \left(n+\frac{1}{2}\right) \chi}
\end{align}
where $C_n = \left(\frac{2}{\pi }\right)^{1/4} \sqrt{\frac{1}{2^n n!}}$ and $H_n$ is a Hermite polynomial of order $n$. These can then be used in the overlap integral \eqref{overlapIntegral} to generate the elements of the matrices shown in Appendix \ref{BasisAppendix}, for example
\begin{equation}
    \braket{\textrm{LG}^1_0 | \textrm{HG}_{20}} = \int_{-\infty}^\infty\int_{-\infty}^\infty \left(\textrm{LG}^1_0\right)^*    \textrm{HG}_{20} \text{d}x\text{d}y = -\frac{1}{\sqrt{2}}
\end{equation}
which is the entry in the first row, second column of the matrix shown in \ref{secondOrderBLGtoHGMatrix}, and appears as the coefficient of the second term in Eq.~\eqref{secondOrderHG}.

\bibliographystyle{unsrt}
\bibliography{bibliography.bib}

\end{document}